\documentclass[conference, 10pt]{IEEEtran}
\IEEEoverridecommandlockouts 
\usepackage[]{graphicx}
\graphicspath{{Figures/}}
\usepackage{caption}
\usepackage{subfig} 
\usepackage{amsmath}
\usepackage{amssymb}
\usepackage{epsfig}
\usepackage{cite}
\usepackage{color}
\usepackage{balance}
\usepackage{amsmath}
\usepackage{amsfonts} 
\usepackage{graphicx}
\usepackage{pdfpages}
\usepackage[T1]{fontenc} 
\usepackage{array}
\usepackage{url}

\usepackage{color}
\usepackage{wrapfig}
\usepackage{lipsum}
\usepackage[hidelinks]{hyperref}
\usepackage{multirow}
\usepackage{tabularx}
\usepackage{tikz}
\usepackage{nth} 
\usepackage{algpseudocode} 
\usepackage{algorithm,algpseudocode}
\usepackage{breqn}
\usepackage{todonotes}
\usepackage{verbatim}

\begin{document}
\title{Embedding Cyclical Information in \\Solar Irradiance Forecasting}
\author{\IEEEauthorblockN{
T.\ A.\ Fathima\IEEEauthorrefmark{1},
Vasudevan~Nedumpozhimana\IEEEauthorrefmark{2},
Yee Hui Lee\IEEEauthorrefmark{3}, and
Soumyabrata Dev\IEEEauthorrefmark{2}\IEEEauthorrefmark{4}
}
\IEEEauthorblockA{\IEEEauthorrefmark{1} Independent Researcher, Dublin, Ireland}
\IEEEauthorblockA{\IEEEauthorrefmark{2} ADAPT SFI Research Centre, Dublin, Ireland}
\IEEEauthorblockA{\IEEEauthorrefmark{3} School of Electrical and Electronic Engineering, Nanyang Technological University (NTU), Singapore}
\IEEEauthorblockA{\IEEEauthorrefmark{4} School of Computer Science, University College Dublin, Ireland}
\thanks{The ADAPT Centre for Digital Content Technology is funded under the SFI Research Centres Programme (Grant 13/RC/2106\_P2) and is co-funded under the European Regional Development Fund.}
\thanks{Send correspondence to S.\ Dev, E-mail: soumyabrata.dev@ucd.ie.}
\vspace{-0.6cm}
}

\maketitle

\begin{abstract}
In this paper, we demonstrate the importance of embedding temporal information for an accurate prediction of solar irradiance. We have used two sets of models for forecasting solar irradiance. The first one uses only time series data of solar irradiance for predicting future values. The second one uses the historical solar irradiance values, together with the corresponding timestamps. 
We employ data from the weather station located at Nanyang Technological University (NTU) Singapore. The solar irradiance values are recorded with a temporal resolution of $1$ minute, for a period of $1$ year. 
We use Multilayer Perceptron Regression (MLP) technique for forecasting solar irradiance. 
We obtained significant better prediction accuracy when the time stamp information is embedded in the forecasting framework, as compared to solely using historical solar irradiance values. 
\end{abstract}

\IEEEpeerreviewmaketitle

\section{Introduction}
One of the key challenges we are facing is the growing energy demand due to the technological advancements and population explosion which necessitate the excess burning of fossil fuels. Besides the higher depletion rate of fossil fuels, these excess burning emits huge amounts of greenhouse gases which causes air pollution and global warming. A reliable, renewable energy source like solar energy is considered as a promising  sustainable source for managing the long term issues in the energy crisis. Solar energy plays an important role in combating air pollution by generating power using solar panels  
and eliminating the emission of harmful greenhouse gases.  Solar energy is one of the promising renewable energy sources for power production based on the above aspects to meet our future energy demands. The annual solar energy output of a photovoltaic system depends on the annual average solar radiation which has to be estimated in advance~\cite{fathima2019chaotic}. However, the uncertainties in the weather patterns severely affect the efficiency of such renewable power production~\cite{Ghazi2014}. A few cloudy, rainy days can have adverse effects on the efficiency of photovoltaic systems as the efficiency of power production is highly dependent on the weather conditions~\cite{dev2019estimating}. The forecasting of solar irradiance (energy received by earth from the sun in the form of electromagnetic radiation) helps to estimate the annual average solar radiation~\cite{fathima2019predicting}. This in turn helps in the smooth operation of switching circuits of the grid connected photovoltaic cells. 

The present work is aimed to develop a  forecasting model of solar irradiance using historical time series data of solar irradiance. We exploit the period nature of solar irradiance value in our forecasting framework. 
This paper is intended to explore the question -- should the cyclical temporal information be embedded in solar irradiance forecasting. 
We use the Multilayer Perceptron Technique (MLP)\cite{Hicham2013} 
for forecasting solar irradiance using one year of historical data. 

The main contributions of this paper include:
\begin{itemize}
    \item we propose a data-driven model for the accurate forecasting of solar irradiance;
    \item we establish the importance of embedding temporal cyclical importance in forecasting model; 
    \item we also share the source-code of our methodology in the spirit of reproducible research\footnote{The code related to this paper is available here: \url{https://github.com/FathimaTA/Forecasting}.}.
\end{itemize}

\section{Solar Irradiance Forecasting}
In this section, we propose two solar irradiance forecasting models by using Multi-Layer Perceptron (MLP) regression method. The first model solely uses the time series data for forecasting solar irradiance. In this model we will embed previous $d$ values of solar irradiance as a $d$ dimensional feature vector and use an MLP regression model to predict the solar irradiance of the next time step. To predict $x_t$ (the solar irradiance at time $t$), the model create a $d$ dimensional feature vector $(x_{t-d},x_{t-d+1}, \cdots x_{t-1})$ from the time series data $(x_1,x_2, \cdots x_t)$ and learn the mapping from the feature vector to the target value $x_t$.

In our second forecasting model, we utilize the cyclic nature of solar irradiance by explicitly using the time stamp information available in the time series data. This temporal information is used as a feature along with previous $d$ time series values of solar irradiance. The solar irradiance is low during nighttime and high during day-time. Therefore, the solar irradiance value exhibits a 
implicit cyclic property with periodicity of $24$ hours. 
Inspired from the work of Szulczyk and Sadique  \cite{Szulczyk2018}, we transformed time stamp values into a two dimensional vector where the first dimension is the \textit{sine} values and the second dimension is the \textit{cosine} values of time stamp. The transformation function for time stamp is described as:


\begin{equation} 
\label{eq1}
T: t \longrightarrow \bigg( sin\big(\frac{2\pi}{\omega}t\big),cos\big(\frac{2\pi}{\omega}t\big)\bigg)
\end{equation}

where $\omega$ is the periodicity, \textit{i.e.}, 24 hours (24*60*60 seconds)

In the proposed MLP based model with time stamp information (tMLP), we transform the time stamp of the solar irradiance value to be predicted by using the above mentioned transformation operator. Subsequently, and we use these two dimensional feature vector (\textit{i.e.}, $T(ts)$ where $ts$ is the time stamp of the value to be predicted) along with $d$ dimensional embedding vector. 
It means that previous $d$ time step values are used to predict the next time step value. We used MLP regression model to predict the next time step value of solar irradiance from the $d+2$ dimensional feature vector ($d$ previous time steps along with the two dimensional transformed time stamp information of the value to be predicted). Therefore, to predict $x_t$ (the solar irradiance at time $t$), the model create a $d+2$ dimensional feature vector $(x_{t-d},x_{t-d+1}, \cdots x_{t-1}, sin\big(\frac{2\pi}{\omega}ts_t\big),cos\big(\frac{2\pi}{\omega}ts_t\big))$ and learn the mapping from the feature vector to the target value $x_t$, where $ts_t$ is the time stamp information at time step $t$.

\section{Results and Discussion}

\subsection{Dataset}
We used 
solar irradiance data with $1$ minute temporal resolution collected from the weather station located at Nanyang Technological University (NTU) Singapore. Such data were recorded over a period of 1 year. In order to understand the efficacy for various lead times, we sampled the original high-resolution data into several datasets with reduced temporal resolutions. We prepared three distinct datasets with varying temporal resolutions of $60$ minutes, $30$ minutes and $15$ minutes respectively. These datasets are referred as Solar$_{60}$, Solar$_{30}$, and Solar$_{15}$ respectively. These datasets contain the averaged solar irradiance data computed across $60$, $30$, $15$ minutes of temporal slices. 



\subsection{Performance evaluation}
We trained MLP based solar irradiance forecasting model 
using historical solar irradiance data, for varying historical period 
$d$ varying from $1$ to $9$. The model is trained on both scenarios -- solely using solar irradiance values (referred as MLP); and one that uses solar irradiance values together with the corresponding timestamps (referred as tMLP). 
We used five-fold cross validation for evaluating the model performance \cite{Berrar2018} of MLP and tMLP, for the different sub-sampled datasets.  We measure the coefficient of determination ($R^2$) value between the actual solar irradiance and forecasted solar irradiance to evaluate model performance. Table~\ref{tab:results} illustrates the results.



\begin{table}[htb]
    \centering
    \caption{We compute the $R^2$ value with- and without- cyclic time feature for varying number of previous time series data ($d$) and varying temporal resolution.}
    \begin{tabular}{c || c | c || c | c || c | c}
    \hline
    &   \multicolumn{2}{c|}{\textbf{Solar$_{60}$}} & \multicolumn{2}{c|}{\textbf{Solar$_{30}$}} & \multicolumn{2}{c}{\textbf{Solar$_{15}$}} \\
    \hline
$d$	&	MLP	&	tMLP	&	MLP	&	tMLP	&	MLP	&	tMLP	\\
\hline
1	&	0.63	&	0.73	&	0.74	&	0.77	&	0.80	&	0.82	\\
2	&	0.65	&	0.73	&	0.74	&	0.78	&	0.81	&	0.82	\\
3	&	0.65	&	0.73	&	0.74	&	0.78	&	0.81	&	0.82	\\
4	&	0.66	&	0.73	&	0.74	&	0.78	&	0.81	&	0.83	\\
5	&	0.67	&	0.72	&	0.75	&	0.78	&	0.80	&	0.82	\\
6	&	0.68	&	0.72	&	0.74	&	0.77	&	0.81	&	0.82	\\
7	&	0.69	&	0.72	&	0.75	&	0.77	&	0.80	&	0.82	\\
8	&	0.70	&	0.72	&	0.75	&	0.77	&	0.80	&	0.82	\\
9	&	0.70	&	0.72	&	0.75	&	0.78	&	0.80	&	0.82	\\
    \hline
    \end{tabular}
    \label{tab:results}
\end{table}

\subsection{Discussions}
Table~\ref{tab:results} illustrates that the forecasting model using time stamp outperforms the one without time stamp. We obtain ~$3$\% increase in the $R^2$ value 
for the three data sets with varying resolutions. Therefore, we conclude 
that periodic nature of solar irradiance is an important information during forecasting. Furthermore, it is observed that higher the resolution of data used in our forecasting model, the better is the model performance. 
We observe considerable 
improvement in the performance, as we gradually increase the number of dimensions from $1$ to $9$. However, we note that there is no significant improvement beyond $6$ dimensions. 

\section{Conclusion \& Future Work}
In this paper, we demonstrate the importance of embedding temporal information for solar irradiance forecasting. We observe that including the temporal feature increases the forecasting accuracy for different historical training period. In the future, we are interested in identifying the impact of the number of previous data (number of dimensions) on the model's performance. 
We are planning to do further work on how to fix the dimensions in our future work. We also plan to use 
deep neural nets~\cite{jain2020forecasting} for better forecasting accuracy with a fixed embedded dimension.

\balance 

\bibliographystyle{IEEEtran.bst}

\end{document}